\documentclass[12pt,preprint]{aastex}
\usepackage{graphicx,color}
\usepackage{natbib}
\usepackage{soul}
\shorttitle{The Transition Corona}
\shortauthors{Masson et al. }

\begin{document}

\title{Dynamics of the Transition corona}

\author{Sophie Masson}
 \affil{Space Weather Laboratory - NASA Goddard Space Flight Center, 
 8800 Greenbelt Road, 
 Greenbelt MD} 

\email{sophie.masson@nasa.gov} 

\and 

\author{Patrick McCauley}
 \affil{Harvard-Smithsonian Center for Astrophysics, 60 Garden St, Cambridge, MA} 
                  
 \and

\author{Leon Golub}
 \affil{Harvard-Smithsonian Center for Astrophysics, 60 Garden St, Cambridge, MA} 
                  
 \and

\author{Katharine K. Reeves}
 \affil{Harvard-Smithsonian Center for Astrophysics, 60 Garden St, Cambridge, MA} 
                  
 \and

\author{Edward E. DeLuca}
 \affil{Harvard-Smithsonian Center for Astrophysics, 60 Garden St, Cambridge, MA}

\begin{abstract}

Magnetic reconnection between open and closed magnetic field in the corona is believed to play a crucial role in the corona / heliosphere coupling. At large scale, the exchange of open /closed connectivity is expected to occur in pseudo-streamer structures. However, there is neither clear observational evidence of how such coupling occurs in pseudo-streamers, nor evidence for  how the magnetic reconnection evolves.

 Using a newly-developed technique, we enhance the off-limb magnetic fine structures observed with AIA and identify a pseudo-streamer-like feature located close to the northern coronal hole. We first identify that the magnetic topology associated with the observation is a pseudo-streamer, null-point related topology bounded by open field. 
 By comparing the magnetic field configuration with the EUV emission regions, we determined that most of the magnetic flux associated with plasma emission are small loops below the pseudo-streamer basic null-point and open field bounding the pseudo-streamer topology. In order to interpret the evolution of the pseudo-streamer, we referred to a 3D MHD interchange reconnection modeling the exchange of connectivity between small closed loops and open field. The observed pseudo-streamer fine structures follow the dynamics of the magnetic field before and after reconnecting at the null-point obtained by  the interchange model. Moreover, the pattern of the EUV plasma emission is the same than the shape of the expected plasma emission location derived from the simulation.

These morphological and dynamical similarities between the pseudo-streamer observations and the results from the simulation strongly suggest that the evolution of the pseudo-streamer, and in particular the opening/closing of the field occurs via interchange/slipping reconnection at the basic null point of the pseudo-streamer. Besides identifying the mechanism at work in the  large-scale coupling between open and closed field, our results highlight that interchange reconnection in pseudo-streamers is a gradual physical process that differs from the impulsive reconnection of the solar-jet model.

\end{abstract}



\section{Introduction}
\label{introduction} 

Understanding how the corona couples with the heliosphere is critical to make advances on the solar wind origin.
Although it is generally admitted that the fast wind originates from the coronal holes, the slow wind origin is still under debate. While \cite{WangSheeley91} proposed that the slow wind comes from the edge of the coronal hole, \cite{Zhao_al09} suggested that interchange reconnection between open and closed field is responsible for the slow wind. The recent S-Web model by \cite{Antiochos_al11} and \cite{Titov_al11} completed this theory by showing that a web of separatrices/quasi-separatrix layers (QSLs), connecting the closed and the open field, is located around the heliospheric current sheet and provides the environment for open/closed connectivity exchange.

The structures ensuring the transition between the open and closed field in the corona are of two kinds. The helmet streamer separates open field of opposite polarity, while the pseudo-streamer (PS) separates open field of  same polarity and therefore includes a null point (NP) at the transition between open and closed field \citep{Wang_al07b,Titov_al12}.
In helmet streamers, the release of material into the heliosphere may result from magnetic reconnection either along the open/closed separatrix surface or at the apex of the streamer, where the field is extended under the solar wind pressure \citep{Wang_al00}. This latter process leads to the formation of plasma blobs that escape into the Heliosphere \citep{Sheeley_al97,Wang_al98}. For opening the field in a PS, it is commonly admitted that interchange reconnection occurs at the null point, exchanging the connectivity between open and closed field \citep{Antiochos_al02,Edmondson_al09,Masson_al12a}. While plasma blobs detaching from the streamer's apex are detected in white-light coronograph images, the dynamics of pseudo-streamers show a quasi-steady state without any distinguishable features above 2 solar radii \citep{WangY_al12}. 
By nature, a pseudo-streamer's cusp is located lower in the corona and consequently, the transition between open and closed field is not detected by coronagraphs, as is the case for streamers. However, several studies showed that the base of a PS formed by two closed lobes bounded by diverging open field can be observed in the low corona at EUV wavelengths \citep{Filippov_al13,Seaton_al13}.  

Even though interchange reconnection is expected in such pseudo-streamer configurations, there is no indication that interchange reconnection is indeed responsible for the observed PS signature or for the field opening. First, while interchange reconnection is usually attributed to a null-point topology surrounded by open field, PSs have in reality a more complex topology formed by several null points linked by separators \citep[see ][]{Titov_al12}. Therefore, the dynamics of the reconnection coupling open and closed field in PSs might not be as straightforward as it is for a single NP. 
Second, previous studies of interchange reconnection have been mostly focused on explaining solar jets \citep{Moreno_al08,Pariat_al09}. Therefore, those interchange models for jets lead to explosive dynamics that are consistent with jet observations \citep{Cirtain_al07}, but they do not correspond to the gradual and quasi-steady evolution of the PS structures at high coronal altitudes \citep{Wang_al07b}.

In this article, we examine the dynamics of EUV observations of a pseudo-streamer-like structure identified in Atmospheric Imaging Assembly (AIA; \cite{Lemen_al12}) data. The high cadence and spatial resolution of AIA allow us to infer the dynamics of the coronal magnetic structures (\S~\ref{AIAobservations}). In section ~\ref{topology}, we present the topology related to the EUV pseudo-streamer and the reconnection regimes that can potentially take place in such PS configuration. 
In order to determine the mechanisms responsible for the PS activity, we first identify the magnetic configuration  where EUV emission is observed (\S~\ref{topological interpretation}). Then, we use results from a 3D MHD simulation of NP interchange reconnection to determine the reconnection regimes that may be at work in the PS open/closed coupling (\S~\ref{mhd model}). Finally, we conclude and discuss our results in \S~\ref{discussion}.

 \section{EUV observations of pseudo-streamer with AIA}
 \label{AIAobservations}
 
 The observations reported here were taken on January 19, 2012 by the AIA instrument aboard the Solar Dynamics Observatory (SDO; \citealt{Pesnell_al12}). We examine a streamer structure on the west limb, at the border of a northern coronal hole located between $35^\circ-60^\circ$~N and between $60^\circ \ge 90^\circ$~W. AIA records full-sun images in seven extreme ultraviolet channels with a 12 s cadence and a resolution of 0.6\arcsec{} per pixel. Of these, the 171 \AA{} band is most suited to our model comparison because it best resolves the fine magnetic structures in the corona. This channel is dominated by Fe IX, with a characteristic temperature of 10\textsuperscript{5.8} \citep{ODwyer_al10}. 

To further enhance fine structure, we have processed the observations using a radial filter. This begins by summing the off-limb component of several images to increase signal-to-noise. The corona beyond the disk is then divided into concentric rings, each of which is scaled as a function of its radius, average brightness, and intensity relative to neighboring rings. As such, flux is not conserved; the brightness of each pixel corresponds only to its intensity relative to other pixels of the same radius (see the SolarSoft routine $<$aia\_rfilter$>$ and http://aia.cfa.harvard.edu/rfilter.shtml for details)
For this application, we have used 10-image sums, yielding images with total integration times of $\sim$20 s over 2 min periods. Figure~\ref{fig1} displays a set of six images between $11:00~\rm{UT}$ and $23:00~\rm{UT}$, and a corresponding movie is available in the online material. While the heated plasma is constrained by the magnetic field, the bright strands observed in AIA indicate explicitly the geometry and the directivity of the magnetic field. 

In Figure~\ref{fig1}, we observe that some of the strands are closed, forming two lobes enclosing a darker region, and some other strands are open and bound to the two closed lobes. The shape of the open strands shows a diverging pattern, as do the fan-like structures identified in EUV wavelengths in the low corona \citep{Habbal_al11,Seaton_al13}. In addition, at the intersection of the closed and open flux, we notice a thin extended region that becomes brighter and thinner as the system evolves. According to numerical studies, such a structure could correspond to the current sheet forming at the sheared null point \citep{RickardTitov96,Galsgaard_al03,Masson_al09b,Pariat_al09,GalsPont11}. 

While such EUV structures have been clearly identified to be related to a PS configuration, the short term dynamics of those structures have never been observed. The high spatial resolution and high time cadence of AIA allow us to identify individual strands and to study their dynamics, providing the evolution of the reconnecting magnetic flux tubes. In the AIA observations, the general pattern of the event does not change drastically during the entire event, but the strands are highly dynamic (see the online material and \S~\ref{dynamics} for details). According to \cite{Wang_al07b}, PS structures observed in the EUV corona should be the result of magnetic reconnection at the cusp of the PS, {\it i.e.} at the null point, between the open and the closed field.



\section{Pseudo-streamer topology and dynamics}
\label{topology}

\subsection{Magnetic field extrapolation}
\label{extrapolation}

On the Carrington map CR2119 observed by HMI /SDO \citep{Schou_al12}, the photospheric magnetic field related to the AIA observations lies between  $35^\circ-70^\circ~\rm{N}$ and $70^\circ - 110^\circ~\rm{W}$ of solar center at $12:00~\rm{UT}$ on Jan 19, 2012. It includes the negative polarity of the north pole, one positive polarity from the trailing polarity of a western decayed active region, and one negative polarity corresponding to the leading polarity of a following decayed active region. 
We perform a Potential Field Source Surface (PFSS) extrapolation \citep{AltschulerNewkirk69} using the SolarSoft routines \citep{SchriDerosa03} and determine the large-scale magnetic topology. Though the PFSS model cannot reproduce the exact magnetic configuration of the corona, it provides a robust method to obtain a reasonable approximation of the global magnetic topology \citep{BrownPriest01}.

Figure~\ref{fig2} shows the magnetic field computed from the PFSS model. The black and pink field lines respectively indicate closed and open field. The closed flux domain can be divided in two categories: small loops ({\it e.g.} orange line) connecting the southern (N1) and the northern (N2) negative polarities to the middle positive polarity (P1), and large loops ({\it e.g.} green line) connecting the N1 and N2  polarities to some eastern positive polarities (P2). Those large closed loops extend up to the source surface and are therefore part of a streamer configuration. For clarity, the large loops connecting N1 to the P2 are not plotted on the Figure~\ref{fig2}, but some of them are plotted on Figure~\ref{fig3}. To the west of the central polarity, the closed flux is confined below open fluxes. Since the fluxes lie at low altitude and are located behind the limb for the AIA field of view, none of them are shown on Figure~\ref{fig2}. Two open magnetic field regions (pink lines) are respectively located north and south of, rather than surrounding, the small closed magnetic flux. In P1 (middle positive polarity), the magnetic connectivity footprints of the low height closed field diverge. The global distribution of the magnetic field is very similar to the topology of one of the pseudo-streamers described by \cite{Titov_al12} (see their Figures 8 and 9).

Even though the PFSS routines does not provide topological analysis, the magnetic field connectivity distribution can be used to derive the main topological elements. In Figure~\ref{fig2}, we draw the separatrices and separators that should be present in a PS topology. We estimate the basic null point location in the region where the open and the closed field converge. By plotting field lines passing in the vicinity of the null point, we determine its approximative location at $r \simeq 1.1~\rm{R}_\odot$ ($47^{\circ}-52.5^{\circ}~\rm{N}$, $95^{\circ}-105^{\circ}~\rm{W}$). From this basic null point emanates a vertical fan surface (or separatrix curtain), shown by the light blue dotted lines, and two spine lines (in red) connecting the two edging negative polarities (N1 and N2). Part of the fan surface is open, connected to the source-surface null line, and part of it is closed, anchored to the photosphere in P2 and P1. Thus, there is a jump in the connectivity mapping of the vertical fan from open to closed flux domain, delimited by an open separator denoted by a dark blue line (Figure~\ref{fig2}). This connects the basic null point and a source surface Y-point located at the cusp of the streamer. A closed separatrix surface,  including the spines, encloses the small loops and separates them from the large loops connected to the P2 polarities. A closed (quasi-)separator line (yellow), passing through the basic null point,  is present at the intersection of this closed separatrix surface and the vertical fan \citep[see][for details]{Titov_al12}. The extremities of this closed (quasi-)separator can be either connected to null points (a true separator) or bald-patches (a quasi-separator).

In addition to the pseudo-streamer separatrices, Quasi-Separatrix Layers (QSLs) are present in the domain  \citep{Antiochos_al11, Titov_al11, Titov_al12} as for a singular null-point topology \citep{Masson_al09a,Masson_al12a}. A quasi-separatrix layer (QSL) defines a region of strong gradients of connectivity \citep{PriestDemoulin95}. Some of them surround the separatrix lines and surfaces (e.g around the closed separatrix surface), and others are present by themselves (e.g. the open QSL formed by open diverging field). QSLs are a preferential site for current formation. Finite-width thin current sheets develop and field lines exchange their connectivity continuously between neighboring field lines \citep{PriHorPon03,Demoulin06,Aulanier_al06}.
The presence of QSLs affect the dynamics of the reconnection during the event and therefore impact the evolution of reconnection signatures \citep[see e.g.][]{Masson_al09b, Aulanier_al07}.

\subsection{Magnetic reconnection for opening magnetic field.}
\label{opening reconnection}

The magnetic topology defines the connectivity of the magnetic field and therefore can be used to predict the global evolution of the magnetic field resulting from reconnection \citep{Demoulin_al94b,Mandrini_al97,Masson_al09b,Reid_al12}.
We use the topology to determine where reconnection should occur and which magnetic fluxes should be involved. 
According to the PS magnetic field configuration, the closed flux can be opened in two different ways.

First, the basic NP delimits the open and the small, closed loops confined below the closed separatrix surface.
When magnetic reconnection occurs at such a null, the closed field loops and the open field exchange their connectivity. This process is the standard interchange reconnection, extensively studied for a regular null-point topology \citep{Pariat_al09,Edmondson_al10,Masson_al12a}, which induces the opening of the closed corona that releases mass and energy into the heliosphere. In the present case, interchange reconnection exchanges the connectivity of the small orange loop with the northern  pink open field line rooted in N2 (see Figure~\ref{fig2}). The initially closed orange line opens and joins the southern open flux (N1). The flux exchange also works the other way, with the northern small closed loops (P1-N2) reconnecting with the southern open field (N1).

The second scenario relies on the association of two reconnection episodes. In addition to the open field and the small closed field, the large loops also converge toward the null point  ({\it e.g.} green line on Figure~\ref{fig2}). Thus, reconnection between the large green and the small orange closed loops occurs at the null point. The initial large green loop closes down below the closed separatrix surface, forming a new small loop connecting the N2 and P1 polarities. The small orange loop jumps outside of the closed separatrix surface and forms a new large loop connecting the N1 and P2 polarities.
While it results in a flux exchange across the vertical fan surface and the closed separatrix surface, it does not open the magnetic field. To do so, the newly reconnected large loop  will open through a classical wind pressure opening or through magnetic reconnection along the open separator. Note that while magnetic reconnection at a closed separator has been studied \citep{Parnell_al10}, magnetic reconnection along an open separator has not. To our knowledge, there is no theoretical work that can help us to determine how magnetic reconnection proceeds along an open separator.

In addition to the regular  interchange reconnection regime, the presence of QSLs surrounding separatrices affects the dynamics of the reconnection signatures during the event. Indeed, a slipping reconnection regime corresponds to a continuous reconnection between neighboring field lines and may be a source of continuous energy release \citep{Baker_al09,Aulanier_al07,Reid_al12,vanDriel_al12}. Thus, this aspect cannot be ignored and the importance of QSL-reconnection for understanding the dynamics of the corona has been demonstrated for flares \citep{Masson_al09b} and CMEs \citep{Savcheva_al12}.
 

\section{Interpretation of the AIA observations }
\label{comparison}

\subsection{Magnetic structures related to the EUV emission}
\label{topological interpretation}


During magnetic reconnection, the energy release leads to heating episodes. Although very hot plasma can be observed during flares, the comparatively gentle and continuous reconnection process suggested by the gradual evolution of the EUV structures (\S~\ref{AIAobservations}) may only heat the plasma to lower temperatures of $\simeq 1~\rm{MK}$. The plasma distributed along the reconnected field then cools and becomes visible in the 171 \AA{} channel, which is sensitive to material at log(T) $\sim$ 5.8 K.


By comparing the magnetic field distribution with the location of the EUV emission from AIA observations, we determine the magnetic flux associated with the emitting plasma and therefore identify the magnetic field that have reconnected. To do so, we select field lines that pass close to the null point, where reconnection should occur, leading to energy release and probably heating episodes.
On Figure~\ref{fig3}, we over plotted these field lines on the AIA image at $15:28~\rm{UT}$, using the same color code than on Figure~\ref{fig2}. The PFSS model allows us to determine the geometry of the field, but it cannot reproduce the exact magnetic configuration. First, the use of a potential field approximation does not contain currents. However, the field is evolving (as suggested by the dynamics of the bright AIA structures) and is, by definition, not potential. Second, the PFSS extrapolation is performed using a synoptic map, and the input photospheric magnetic field has been measured 3 days before the event when the region was at the disk center. The coronal activity and the photospheric motions have therefore changed the magnetic field distribution, but the global topology resulting from the large scale photospheric magnetic field is expected to be similar.


Even though the match is not perfect, 
we can still identify the magnetic flux associated with the EUV structures.
The small loops are related to the bright closed structures confining a dark cavity, which we identified as a closed magnetic flux below the null point in the AIA observations. The open flux and the large loops anchored in N1 and N2 polarities are localized where the AIA open bright strands are observed.
 Furthermore, the null-point location is close to the thin and intense bright elongated spot (see \S\ref{AIAobservations}), supporting that the bright sheet corresponds to the null-point current sheet. 

The association between the EUV structures and the magnetic fluxes, as well as the null-point current sheet, provides a consistent picture supporting that the bright EUV structures are at least partially caused by null-point reconnection between closed flux below the null and the flux outside of it. As we described in \S~\ref{opening reconnection}, the opening can be achieved through a direct interchange reconnection, but also through a two step process which first implies closed/closed reconnection between small and large loops, followed by an open/closed reconnection along the open separator. 

In order to observe an increase of plasma emission,  magnetic flux tubes have to be denser than their surrounding medium. 
Magnetic reconnection provides mechanisms that can trigger the filling of loops by plasma material \citep[see][references therein]{Baker_al09}, such as chromospheric evaporation \citep{DelZanna08} and rarefaction wave driven by interchange reconnection between tenuous open and dense closed magnetic flux \citep{Bradshaw_al11}. Whatever the process is, the time needed to fill a coronal loop is on the order of tens of minutes at most  \citep{Reeves_al07}. This is much shorter than the duration of the whole event, which lasts several hours. Therefore, if closed/closed reconnection occurs between the large and small loops, we expect to observe plasma emission all along the new reconnected loops anchored in P2 polarity, rather than only in the region co-spatial with the open flux. On Figure~\ref{fig3}, as well as in animations 2 \& 3, it clearly appears that no EUV emission is observed along the large closed field, suggesting that the AIA bright open strands are related to the open flux. Thus, EUV emission is mostly distributed along the closed flux below the null point and the open flux, suggesting that interchange reconnection between open and closed flux contributes significantly to the dynamics of the pseudo-streamer.




\subsection{A 3D model of interchange reconnection}
\label{mhd model}

In the following, we present the results of a 3D MHD simulation from \cite{Masson_al12a}, hereafter M12, of an interchange reconnection in order to determine whether such open/closed reconnection can explain the observational conjectures in the previous section. Although this modeling is based on a NP, the PS topology has the same topological objects involved in null-point interchange reconnection: the null point, the closed separatrix surface and the open QSLs.
By comparing the numerical results for a single null-point topology and the PS observations, we expect to provide some clues to understand the dynamics of the event. We are aware that the dynamics of reconnection for null-points and pseudo-streamers should not be exactly the same, but we believe that the interchange at the null-point should behave quite similarly. Besides, to our knowledge, there are no studies of the dynamics in pseudo-streamers, and the 3D null-point reconnection model is the most sophisticated model that we have to simulate interchange reconnection at a null point.
 %
For our particular example, as we suggested in \S~\ref{topological interpretation}, null-point reconnection between open and closed field seems to be predominant and single null-point dynamics may suffice to capture the essential aspects of the open/closed coupling PS dynamics.

\subsubsection{Dynamics of the magnetic field}
\label{dynamics}

 The tridimensional MHD simulation of M12 modeled an asymmetric null-point topology, with a dome-like fan surface surrounded by an open diverging magnetic field, and studied the dynamics of interchange reconnection \citep[see][for details]{Masson_al12a}.
The initial atmosphere is stratified, obeying hydrostatic equilibrium, and the temperature, gravity and density profiles are adjusted to obtain a solar-like regime: $\beta$ is lower than 1 in the corona except at the null-point where $\beta$ becomes greater than $1$. 
In order to emulate the physics of the solar corona, the top and side boundaries of the numerical box are
open, whereas the photospheric boundary is reflecting and line-tied. 
The system is forced by applying a sub-Alvf\'enic photospheric velocity flow, which moves the positive polarity toward negative y in the area of $x \in [-7, 3]$ and $y \in [10, -30]$, where a part of the fan is rooted. 

This flow shears only a part of the closed dome-like fan surface, increasing the asymmetry of the NP topology along the $y$-direction and leading to the extension of the open QSL-halo around the spine (Figure 5 in \cite{Masson_al12a}). It also leads to the bulging of field lines that compress the separatrices and a region of over-density forms along the separatrices (Figure~\ref{fig4} and Figure 2 in \cite{Masson_al12a}). Moreover, it induces a compression of the null-point and misaligns the inner and the outer spine.
Magnetic reconnection is expected to occur at the null point in order to bring the spines back into alignment \citep{RickardTitov96,Antiochos_al02,Galsgaard_al03}.

In order to determine the dynamics of magnetic reconnection, \cite{Masson_al12a} plotted field lines from footpoints fixed in the advected positive polarity, integrated them up to their conjugate footpoints, and followed the evolution of the connectivities during the simulation. 
The top row of Figure~\ref{fig4} shows the evolution at 3 different times  during the simulation. Initially, 3 groups of colored field lines have been plotted closed below the fan, and 4 groups of colored field lines open into the corona. The field lines are plotted from fixed footpoints. Following the connectivities of the conjugate footpoints of those field lines, M12  shows that magnetic reconnection occurs at the null-point through interchange reconnection. The initially closed and open magnetic field lines exchange their connectivity and respectively open into the corona and close down below the fan surface.
In addition, the field lines show an apparent slipping motion before and after the null-point interchange reconnection. These dynamics result from the slipping reconnection regime across the open QSL that surround the fan-spine topology (\cite{Masson_al12a} see online material, animation1). 

 The AIA observations show a 2D projected evolution of a purely 3D phenomenon. It is therefore difficult to accurately follow the dynamics of the bright structures. 
 We notice that some initially closed field lines grow and approach the closed separatrix surface before disappearing (see animation 2). Meanwhile, there are also loops on the upper side of the null point that collapse into the null (animation 2 and 3). Around 14:00 -15:00~\rm{UT} there are loops underneath the null that shrink down, moving in the opposite sense of the loop pointed to in animation 2. Combined with line-of-sight effects, these field line motions are consistent with the dynamics of the interchange reconnection of M12. The initially closed field lines bulge and slip toward the null point, then reconnect at the null point and open in the corona. Furthermore, the open flux tubes move away from the null-point area and show an apparent slipping motion  (see animation 3). Such motions of the strands may correspond to the apparent motion of open field lines induced by the continuous reconnection across the open QSL (diverging open magnetic field), as suggested previously by \cite{Aulanier_al07} for soft X-ray emission.

The dynamics of the fine structures observed with AIA show some strong similarities with the dynamics of the magnetic field derived from the M12 simulation. Although the dynamics of the interchange reconnection in a simple NP topology can not reproduce the full dynamics of the PS evolution, it provides some elements to understand it. Thus, the hybrid regime combining NP-interchange and slipping reconnection seems to reproduce the dynamics of the magnetic field derived from the EUV strands. This scenario leads us to propose that such an interchange/slipping regime may be responsible for the closed/open coupling leading to the energy release into the heliosphere.


\subsubsection{Morphology of the plasma emission regions}
\label{morphology}

In order to support our interpretation of the mechanism powering the PS observation, we verify that such NP-interchange/slipping reconnection in an asymmetric null-point topology can produce radiative signatures of reconnection similar to the PS observation.
%
%
The numerical model of M12 focuses on the magnetic field dynamics but does not treat properly the plasma response. In the continuity equation, a nonphysical explicit diffusive term, which smooths the gradients, has been added. This term helps to stabilize the computation by avoiding cavitation or shock formation. In addition, the energy equation is reduced to the adiabatic temperature equation, to which a nonphysical explicit diffusive term has also been added for gradient smoothing. Therefore, the plasma is only heated by adiabatic effects and cools down via the diffusive term. Any other heating or cooling processes (e. g.  Joule dissipation or radiation losses) are not included in the simulation. Since the density and temperature evolution do not reflect a correct plasma response (\S~\ref{mhd model}), we cannot generate synthetic images of the emission \citep{Reeves_al10} for comparison with the AIA observations. 

Instead, we propose to use the current density distribution from M12 simulation to determine where magnetic reconnection occurs. Given that the reconnection is happening continuously over several hours, we can  estimate the location of cooling plasma that would be observed by the AIA 171 \AA\ channel along the reconnected field.
Intense currents are expected to be localized around the null point and along the separatrices,  but also along the QSLs \citep{Aulanier_al06}. In order to increase the contrast of regions where strong currents are expected, we use $\alpha= {\bf j}. {\bf b}/b^2$. This quantity is less dependent on the magnetic field magnitude but takes into account the orientation of the currents with respect to the field.  Current sheets in the regions of weak magnetic field and strong magnetic gradient distributed on a coarse grid, {\it e.g.} the open QSL surrounding the outer spine, are highlighted by  $\alpha$. 
%
According to the magnetic field distribution (Figure~\ref{fig3}), the line-of-sight of the AIA observations aligns with the direction of the longitudinal extension axis of the closed flux. In order to be consistent with the AIA observations, which are integrated along the line-of-sight, we integrate the $\alpha$ parameter along the $y$ axis (extension axis of the closed flux in M12) from the boundary $y=-30$ to the ({\it x,z})-plane at $y=0$. This integration range avoids the part of the QSL located between $y=0$ and $y=30$, which results from the symmetry of the system \citep[see][]{Masson_al12a}. 

The bottom row of Figure~\ref{fig4} shows the 2D-maps of the integrated value of $\alpha$ at three different times. First, we identify the magnetic structures associated with the $\alpha$ structures. The brightest region indicated on this 2D $\alpha$-map corresponds to the current sheet located around the null point. The arc-shaped white structure corresponds to the fan separatrix (delimiting the open and the closed magnetic field). Around $x=-15$ to $x=-10$,  the black line embedded in the white halo corresponds to the inner and the outer spines (refer to labels on Fig.~\ref{fig4}). Finally, the extended white-halo structure covering the area between $x \in[-15; -5] $ and $y \in [0;20]$ is co-spatial with the reconnected open field lines belonging to the QSL.

%
%
Even though the intensity of the integrated-$\alpha$ does not reproduce the emission, it shows the location where energy release is expected and where plasma emission may be found. Previous studies have already shown an excellent association between the current density from simulations and the hot plasma emission at different wavelengths \citep[e.~g.][]{Delannee_al08,Aulanier_al10,Savcheva_al12}.  
Comparing the AIA images and the synthetic $\alpha$-map, we notice that the emitting regions in AIA display the same patterns as the structures in the 2D-$\alpha$ map. 
We indeed find the thin and bright region at the null location, the arc-shape white structures for the closed field and separatrix surface, and the extended bright open structure related to the open QSL. Similarly, the bright structure dividing the closed separatrix in two is also present in the $\alpha$-map and the AIA observations. We also notice that the bright structure related to the null point on the observations (top-middle and right panel of Figure~\ref{fig4}) and on the $\alpha$-map (bottom row of Figure~\ref{fig4}) becomes thinner and more intense as time goes on.
Thus, the PS EUV emission shows the same reconnection signatures as the interchange reconnection in a single asymmetric null-point topology. Those similarities strongly support our interpretation that the NP-interchange and slipping reconnection regimes are responsible for the PS dynamics observed with AIA.


%

\section{Conclusion and Discussion}
\label{discussion}

We explored the dynamics of a pseudo-streamer structure observed by AIA and suggested what mechanisms are at work by combining the magnetic topology of the event and a 3D MHD model of interchange reconnection from \cite{Masson_al12a}, hereafter M12. 
The AIA 171~\AA\ observations presented in this paper occurred on January 19, 2012 and correspond to an EUV pseudo-streamer. By applying a radial filter, we showed that the fine structures that trace magnetic fluxes are highly dynamic, while the global pseudo-streamer structure is more of a quasi-steady type (\S~\ref{AIAobservations}).
Using the PFSS extrapolation, we showed that the magnetic configuration related to the  AIA observations is a pseudo-streamer topology as defined by \cite{Titov_al12}.   It is now well accepted that the dynamics of magnetic reconnection and the resulting radiative signatures are defined by the topology  \citep{Masson_al09b,Reid_al12,Sun_al13}. 
For coronal field opening in a pseudo-streamer topology, interchange reconnection occurs at the null point either between the small loops and the open field or between small and large loops followed by reconnection along the open separator with the open field (\S~\ref{extrapolation}).
 By comparing the location of the AIA/EUV structures and the magnetic field distribution from the PFSS extrapolation, we showed that interchange reconnection at the null-point between closed and open field seems a plausible explanation of the pseudo-streamer  observations. 
The pseudo-streamer and null-point topology are different, but both have a basic null point where interchange reconnection can occur between open and closed field. We expect to observe a behavior very similar to the null-point topology given the observed pseudo-streamer  topology.
First, we showed that the dynamics of the open and closed magnetic field in the M12 interchange/slipping reconnection model is very similar to the dynamics of the white EUV strands in the AIA observations (\S~\ref{topological interpretation}).
Second, we found that the distribution of plasma emission expected by the M12  model has a shape very similar to the pseudo-streamer  observation, showing that single null-point reconnection, as in M12, can reproduce the observational signature of a pseudo-streamer (\S~\ref{mhd model}).
 
\begin{figure*}
\vspace{1cm}
\centerline{
\includegraphics[width=0.9\textwidth,clip=]{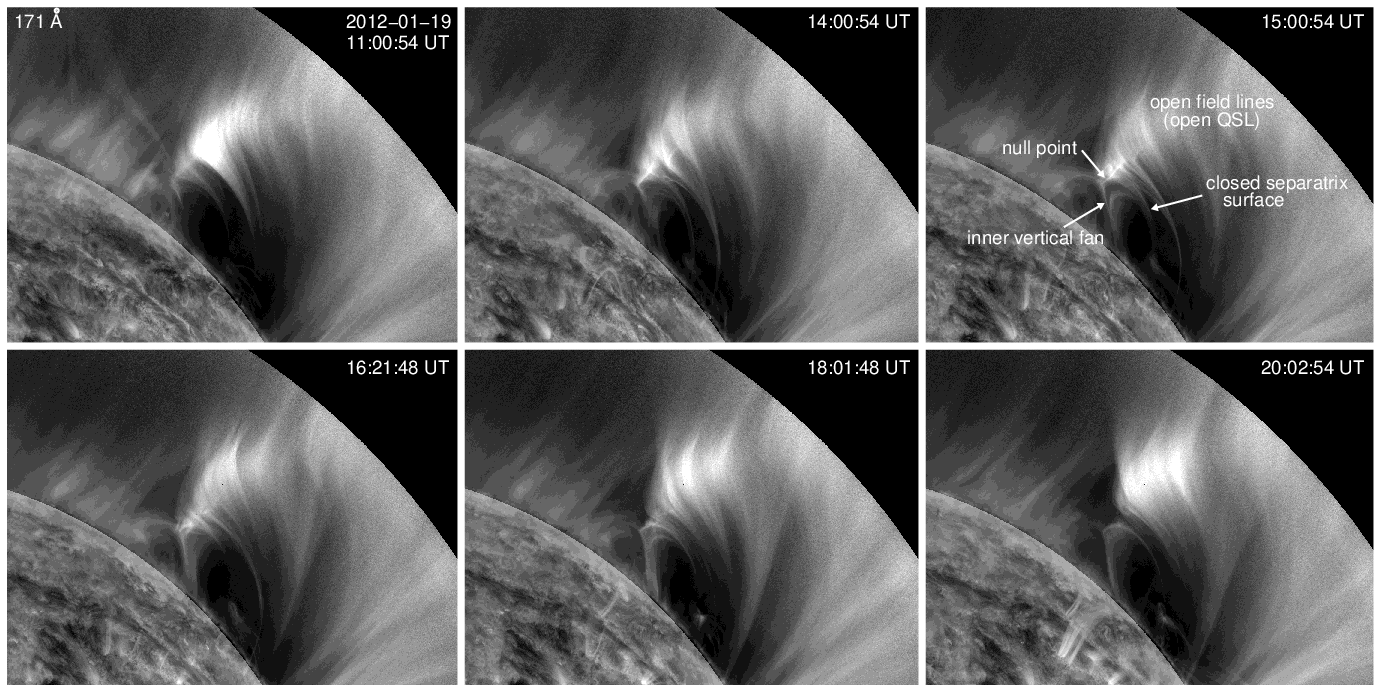}
 }
\caption{Temporal sequence showing the evolution of the emission at 171~{}\AA~observed by AIA / SDO. We distinguish the open and closed  field lines and the null-point current sheet. On the top right panel we labelled this structures accordingly to the pseudo-streamer topology described in \S~\ref{extrapolation}.}
  \label{fig1}
\end{figure*}

\begin{figure}
\vspace{1cm}
\centerline{
\includegraphics[width=0.9\textwidth,bb=0 0 1536 1520,clip=]{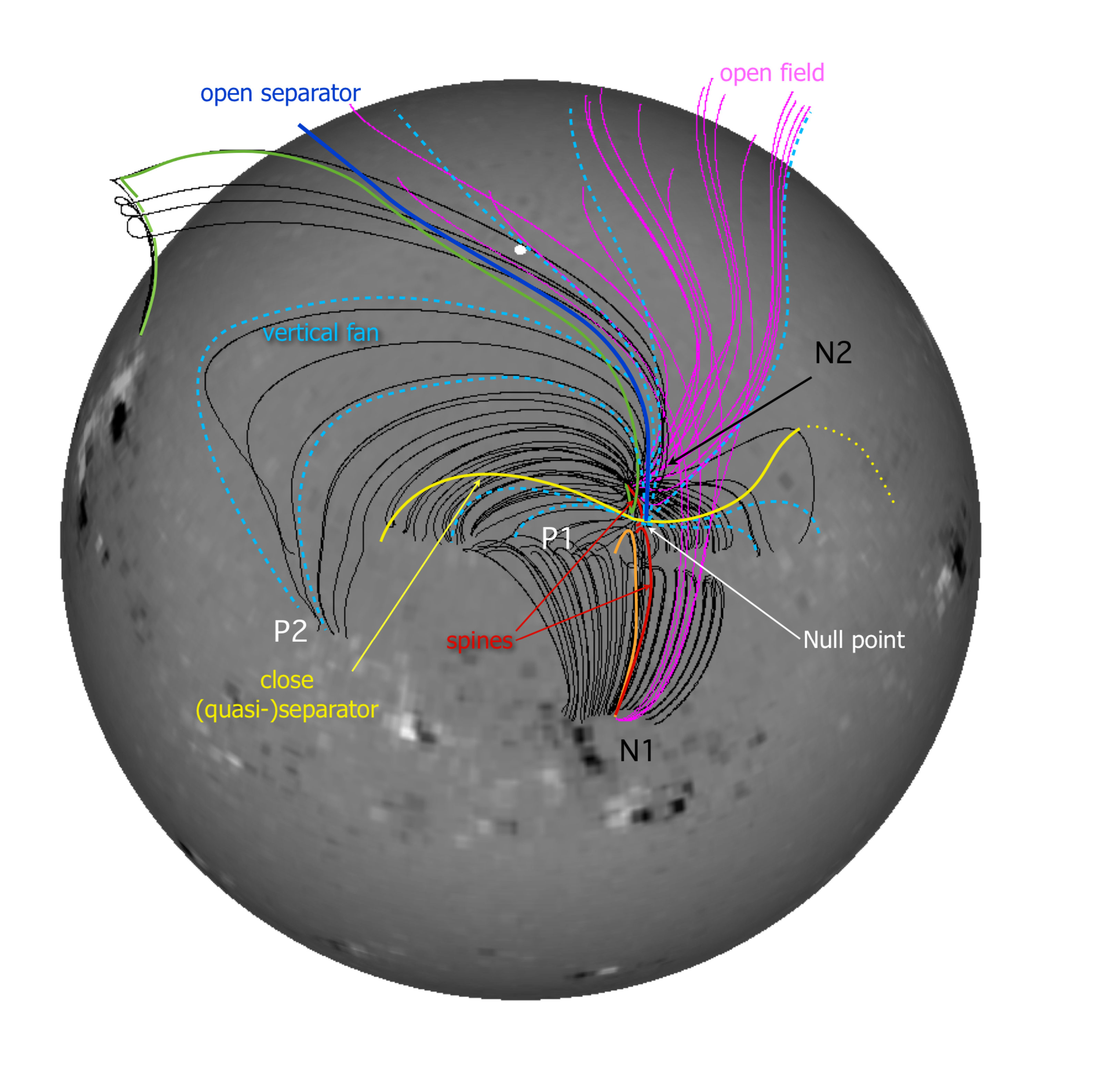}
 }
\caption{Topology of the pseudo-streamer from the PFSS extrapolation. The black field lines show the closed field lines whereas the pink field lines show the open field toward the interplanetary medium. The two red lines show the spine and the light blue dotted lines shows the vertical fan surface. The yellow line passes though the null point shows the closed (quasi-)separator, while the dark blue plain and dotted line corresponds to the two open separator. The green and the orange field lines show respectively the large and small closed loops.}
  \label{fig2}
\end{figure}

\begin{figure}
\vspace{1cm}
\centerline{
\includegraphics[width=0.9\textwidth,bb=0 0 1536 1520,clip=]{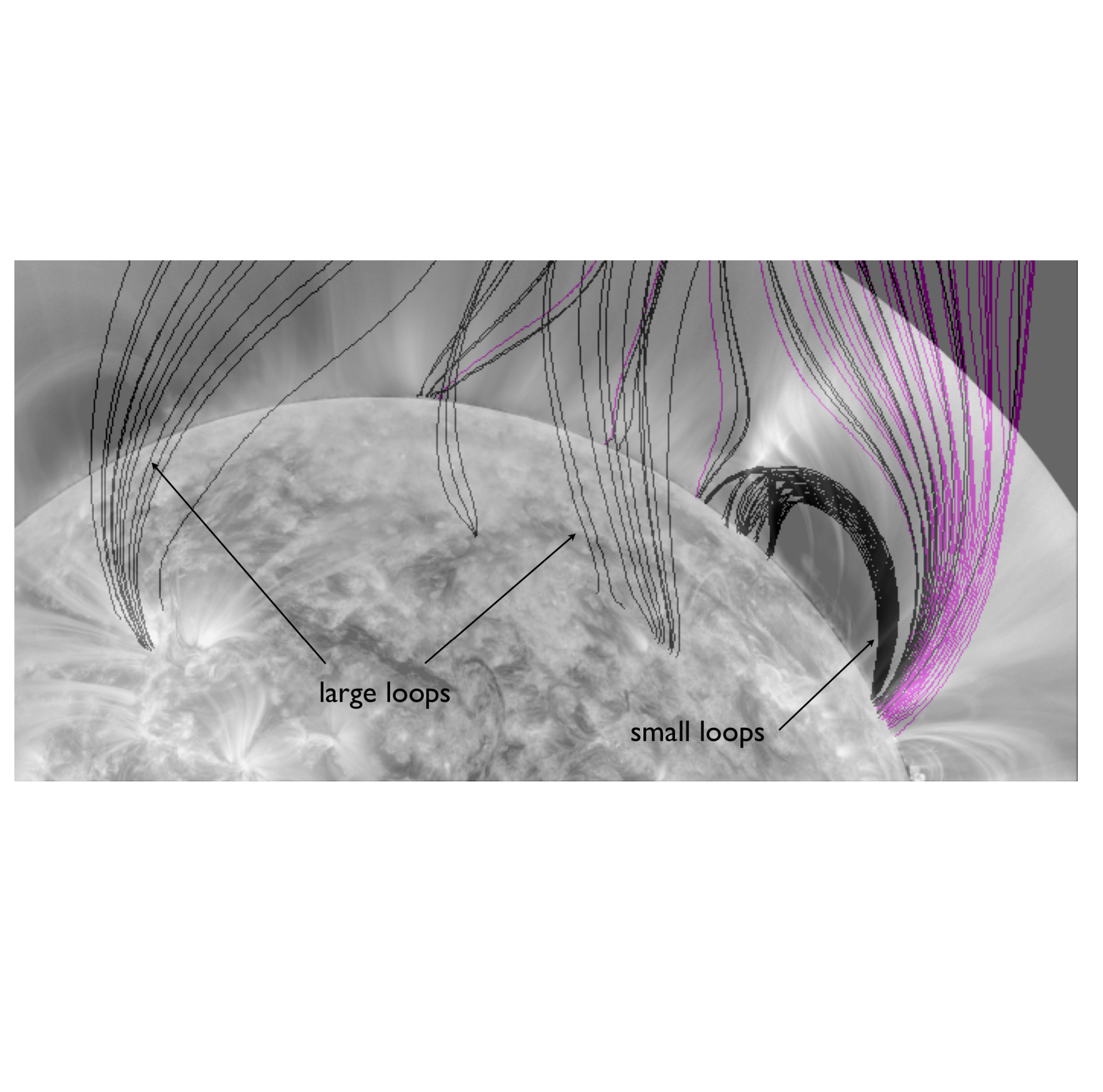}
 }
\caption{Sample of the open and closed field lines passing close to the null point over-plotted on the AIA observations. The EUV emissions are located along the closed separatrix surface and the open field. No EUV emission is observed where the large closed loops are anchored.}
  \label{fig3}
\end{figure}

\begin{figure*}
\centerline{
\vspace{1cm}
\includegraphics[width=0.95\textwidth,clip=]{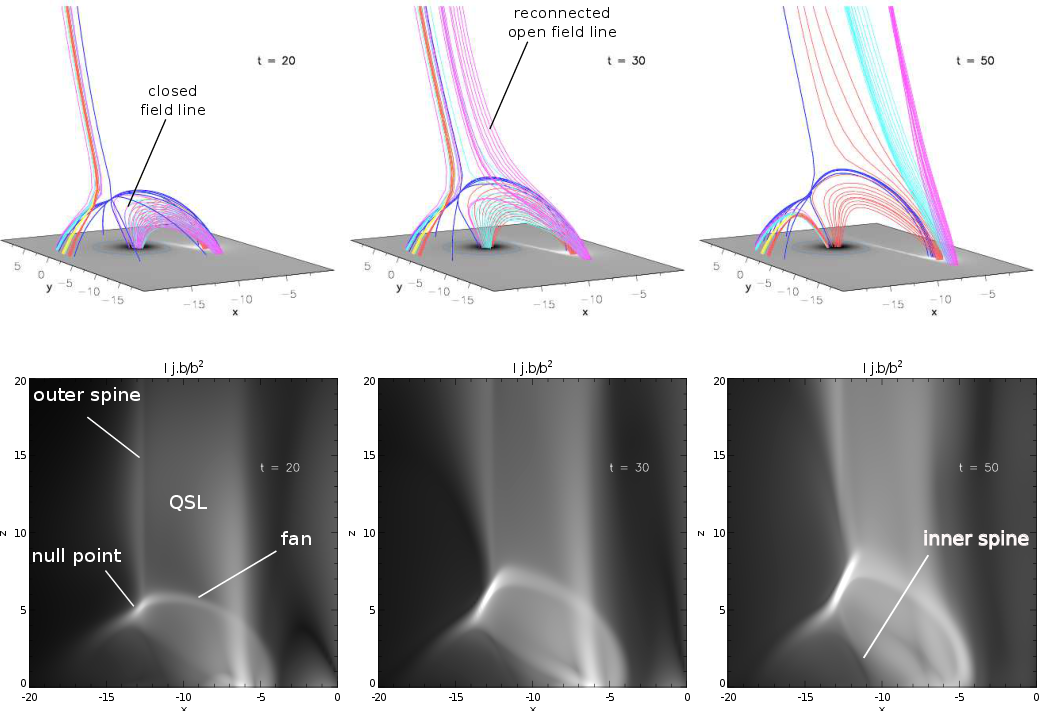}
 }
\caption{Results of the M12 simulation. Top row: Evolution of the magnetic field lines during the interchange reconnection process. The greyscale code at the bottom boundary represents the vertical component of the magnetic field, white and black color show respectively the positive and negative values. The dark blue field lines denote the fan and spine separatrix field lines and the group of colored field lines (pink, red, yellow, light blue) displays the open and the closed field lines that reconnect at the null point. Bottom row:  Temporal series of the 2D map of the integrated value of $\alpha^{0.33}$. The greyscale coding is black for $\alpha =0$ and white for $\alpha = 2.34$ )}.
  \label{fig4}
\end{figure*}

The location of the EUV emission, the similarities of the magnetic field dynamics, and the morphology of the radiative signatures are all pointing to the same conclusion: the pseudo-streamer observations result from the interchange/slipping reconnection at the basic pseudo-streamer null-point.
%
%
It is important to mention that the second scenario (see \S~\ref{opening reconnection}) involving magnetic reconnection at the open separator cannot be ruled out. It may also be involved in the dynamics of the pseudo-streamer, but it does not seem to dominate this particular event. Additional investigations into the dynamics of magnetic reconnection in a pseudo-streamer topology are needed to confirm the role of interchange reconnection at a basic null-point, to determine the additional reconnection regime involved, and to establish how they affect the dynamics.
Nonetheless, this study offers the first observational evidence for interchange reconnection happening in a PS structure, where open and closed field are coupled through reconnection, leading to energy and mass transfer from the corona into the heliosphere. 
In addition, the M12 model deduces that the open white strands (see Fig.~\ref{fig1}) result from the QSL-reconnection, implying a continuous release of energy that can generate plasma emission lasting for several hours \citep[see][for details]{Baker_al09}. This potentially explains why the open fluxes remained bright all along the event, rather than fading gradually after the NP reconnection.
Note that the scaling applied to the model (\S~\ref{mhd model})and the observations (\S~\ref{AIAobservations}) enhances the faint structures at high coronal height. This differential method used to produce the AIA images shown here allows us to better visualize the QSL structures in the corona.





One interesting characteristic of the M12 model is that the interchange reconnection is a gradual process, as in \cite{Edmondson_al09}, which is consistent with the dynamics of the observed pseudo-streamer that lasted for several hours without any drastic changes \citep[see][]{Seaton_al13}. According to \cite{Wang_al00}, pseudo-streamer configurations store and release gradually the energy over several hours and maybe days.
Such behavior contrasts the explosive nature of solar jets \citep{Cirtain_al07} also caused by interchange reconnection at a single null-point \citep{Pariat_al09}. 
Even though both type of events are caused by interchange reconnection at null-point, their dynamics result from two distinct reconnection  regimes. Therefore, the explosive interchange solar-jet model can not be applied to model reconnection dynamics in the large scale pseudo-streamer.

%
 


Although the exact role of the interchange reconnection in the origin of the solar wind is still under debate \citep[see e. g.][]{WangY_al12}, it is very likely that this process is essential to releasing mass and energy into the heliosphere and that it contributes somehow to the solar wind generation.  
 In agreement with \cite{Antiochos_al11,Titov_al11}, the global topology of the solar coronal magnetic field displays a web of similar pseudo-streamer topologies delimiting the open and closed magnetic field. This web of (Quasi-)Separatrices provides the environment for opening the closed coronal field through interchange reconnection all around the Sun.
Moreover, \cite{Seaton_al13} showed that the fan-like structures, with a cusp located low in the corona (below 2 solar radii), are present over the entire solar cycle. Scanning the AIA data, we also find that the type of event presented in this paper is commonplace. This finding suggests that interchange/slipping reconnection processes may occur frequently in the pre-disposed S-Web environment, and according to our results, relies on gradual interchange/slipping reconnection. Such solar phenomena may continuously release a significant amount of mass and energy into the heliosphere \citep{vanDriel_al12} and therefore may play a critical role in the wind generation.



 %

\section{Acknowledgements} The authors thanks the referees for the pertinent comments that helped to improve the manuscript, as well as J. Klimchuck and G. Aulanier for helpful discussions. This work partially supported under contract SP02H1701R from Lockheed-Martin to SAO. SM gratefully acknowledges support from the NASA Postdoctoral Program, administrated by Oak Ridge Associated University through a contract with NASA, during her stay at NASA Goddard Space Flight Center.

\bibliographystyle{apj}
\bibliography{../../phd_biblio}  

\begin{thebibliography}{51}
\expandafter\ifx\csname natexlab\endcsname\relax\def\natexlab#1{#1}\fi

\bibitem[{{Altschuler} \& {Newkirk}(1969)}]{AltschulerNewkirk69}
{Altschuler}, M.~D. \& {Newkirk}, G. 1969, \solphys, 9, 131

\bibitem[{{Antiochos} {et~al.}(2002){Antiochos}, {Karpen}, \&
  {DeVore}}]{Antiochos_al02}
{Antiochos}, S.~K., {Karpen}, J.~T., \& {DeVore}, C.~R. 2002, \apj, 575, 578

\bibitem[{{Antiochos} {et~al.}(2011){Antiochos}, {Miki{\'c}}, {Titov},
  {Lionello}, \& {Linker}}]{Antiochos_al11}
{Antiochos}, S.~K., {Miki{\'c}}, Z., {Titov}, V.~S., {Lionello}, R., \&
  {Linker}, J.~A. 2011, \apj, 731, 112

\bibitem[{{Aulanier} {et~al.}(2007){Aulanier}, {Golub}, {DeLuca}, {Cirtain},
  {Kano}, {Lundquist}, {Narukage}, {Sakao}, \& {Weber}}]{Aulanier_al07}
{Aulanier}, G., {Golub}, L., {DeLuca}, E.~E., {Cirtain}, J.~W., {Kano}, R.,
  {Lundquist}, L.~L., {Narukage}, N., {Sakao}, T., \& {Weber}, M.~A. 2007,
  Science, 318, 1588

\bibitem[{{Aulanier} {et~al.}(2006){Aulanier}, {Pariat}, {D{\'e}moulin}, \&
  {Devore}}]{Aulanier_al06}
{Aulanier}, G., {Pariat}, E., {D{\'e}moulin}, P., \& {Devore}, C.~R. 2006,
  \solphys, 238, 347

\bibitem[{{Aulanier} {et~al.}(2010){Aulanier}, {T{\"o}r{\"o}k}, {D{\'e}moulin},
  \& {DeLuca}}]{Aulanier_al10}
{Aulanier}, G., {T{\"o}r{\"o}k}, T., {D{\'e}moulin}, P., \& {DeLuca}, E.~E.
  2010, \apj, 708, 314

\bibitem[{{Baker} {et~al.}(2009){Baker}, {van Driel-Gesztelyi}, {Mandrini},
  {D{\'e}moulin}, \& {Murray}}]{Baker_al09}
{Baker}, D., {van Driel-Gesztelyi}, L., {Mandrini}, C.~H., {D{\'e}moulin}, P.,
  \& {Murray}, M.~J. 2009, \apj, 705, 926

\bibitem[{{Bradshaw} {et~al.}(2011){Bradshaw}, {Aulanier}, \& {Del
  Zanna}}]{Bradshaw_al11}
{Bradshaw}, S.~J., {Aulanier}, G., \& {Del Zanna}, G. 2011, \apj, 743, 66

\bibitem[{{Brown} \& {Priest}(2001)}]{BrownPriest01}
{Brown}, D.~S. \& {Priest}, E.~R. 2001, \aap, 367, 339

\bibitem[{{Cirtain} {et~al.}(2007){Cirtain}, {Golub}, {Lundquist}, {van
  Ballegooijen}, {Savcheva}, {Shimojo}, {DeLuca}, {Tsuneta}, {Sakao}, {Reeves},
  {Weber}, {Kano}, {Narukage}, \& {Shibasaki}}]{Cirtain_al07}
{Cirtain}, J.~W., {Golub}, L., {Lundquist}, L., {van Ballegooijen}, A.,
  {Savcheva}, A., {Shimojo}, M., {DeLuca}, E., {Tsuneta}, S., {Sakao}, T.,
  {Reeves}, K., {Weber}, M., {Kano}, R., {Narukage}, N., \& {Shibasaki}, K.
  2007, Science, 318, 1580

\bibitem[{{Del Zanna}(2008)}]{DelZanna08}
{Del Zanna}, G. 2008, \aap, 481, L49

\bibitem[{{Delann{\'e}e} {et~al.}(2008){Delann{\'e}e}, {T{\"o}r{\"o}k},
  {Aulanier}, \& {Hochedez}}]{Delannee_al08}
{Delann{\'e}e}, C., {T{\"o}r{\"o}k}, T., {Aulanier}, G., \& {Hochedez}, J.-F.
  2008, \solphys, 247, 123

\bibitem[{{D{\'e}moulin}(2006)}]{Demoulin06}
{D{\'e}moulin}, P. 2006, \asr, 37, 1269

\bibitem[{{D\'emoulin} {et~al.}(1994){D\'emoulin}, {Mandrini}, {Rovira},
  {Henoux}, \& {Machado}}]{Demoulin_al94b}
{D\'emoulin}, P., {Mandrini}, C.~H., {Rovira}, M.~G., {Henoux}, J.~C., \&
  {Machado}, M.~E. 1994, \solphys, 150, 221

\bibitem[{{Edmondson} {et~al.}(2010){Edmondson}, {Antiochos}, {DeVore},
  {Lynch}, \& {Zurbuchen}}]{Edmondson_al10}
{Edmondson}, J.~K., {Antiochos}, S.~K., {DeVore}, C.~R., {Lynch}, B.~J., \&
  {Zurbuchen}, T.~H. 2010, \apj, 714, 517

\bibitem[{{Edmondson} {et~al.}(2009){Edmondson}, {Lynch}, {Antiochos}, {De
  Vore}, \& {Zurbuchen}}]{Edmondson_al09}
{Edmondson}, J.~K., {Lynch}, B.~J., {Antiochos}, S.~K., {De Vore}, C.~R., \&
  {Zurbuchen}, T.~H. 2009, \apj, 707, 1427

\bibitem[{{Filippov} {et~al.}(2013){Filippov}, {Koutchmy}, \&
  {Tavabi}}]{Filippov_al13}
{Filippov}, B., {Koutchmy}, S., \& {Tavabi}, E. 2013, \solphys, 286, 143

\bibitem[{{Galsgaard} \& {Pontin}(2011)}]{GalsPont11}
{Galsgaard}, K. \& {Pontin}, D.~I. 2011, \aap, 534, A2

\bibitem[{{Galsgaard} {et~al.}(2003){Galsgaard}, {Priest}, \&
  {Titov}}]{Galsgaard_al03}
{Galsgaard}, K., {Priest}, E.~R., \& {Titov}, V.~S. 2003, \jgr, 108, 1042

\bibitem[{{Habbal} {et~al.}(2011){Habbal}, {Druckm{\"u}ller}, {Morgan}, {Ding},
  {Johnson}, {Druckm{\"u}llerov{\'a}}, {Daw}, {Arndt}, {Dietzel}, \&
  {Saken}}]{Habbal_al11}
{Habbal}, S.~R., {Druckm{\"u}ller}, M., {Morgan}, H., {Ding}, A., {Johnson},
  J., {Druckm{\"u}llerov{\'a}}, H., {Daw}, A., {Arndt}, M.~B., {Dietzel}, M.,
  \& {Saken}, J. 2011, \apj, 734, 120

\bibitem[{{Lemen} {et~al.}(2012){Lemen}, {Title}, {Akin}, {Boerner}, {Chou},
  {Drake}, {Duncan}, {Edwards}, {Friedlaender}, {Heyman}, {Hurlburt}, {Katz},
  {Kushner}, {Levay}, {Lindgren}, {Mathur}, {McFeaters}, {Mitchell}, {Rehse},
  {Schrijver}, {Springer}, {Stern}, {Tarbell}, {Wuelser}, {Wolfson}, {Yanari},
  {Bookbinder}, {Cheimets}, {Caldwell}, {Deluca}, {Gates}, {Golub}, {Park},
  {Podgorski}, {Bush}, {Scherrer}, {Gummin}, {Smith}, {Auker}, {Jerram},
  {Pool}, {Soufli}, {Windt}, {Beardsley}, {Clapp}, {Lang}, \&
  {Waltham}}]{Lemen_al12}
{Lemen}, J.~R., {Title}, A.~M., {Akin}, D.~J., {Boerner}, P.~F., {Chou}, C.,
  {Drake}, J.~F., {Duncan}, D.~W., {Edwards}, C.~G., {Friedlaender}, F.~M.,
  {Heyman}, G.~F., {Hurlburt}, N.~E., {Katz}, N.~L., {Kushner}, G.~D., {Levay},
  M., {Lindgren}, R.~W., {Mathur}, D.~P., {McFeaters}, E.~L., {Mitchell}, S.,
  {Rehse}, R.~A., {Schrijver}, C.~J., {Springer}, L.~A., {Stern}, R.~A.,
  {Tarbell}, T.~D., {Wuelser}, J.-P., {Wolfson}, C.~J., {Yanari}, C.,
  {Bookbinder}, J.~A., {Cheimets}, P.~N., {Caldwell}, D., {Deluca}, E.~E.,
  {Gates}, R., {Golub}, L., {Park}, S., {Podgorski}, W.~A., {Bush}, R.~I.,
  {Scherrer}, P.~H., {Gummin}, M.~A., {Smith}, P., {Auker}, G., {Jerram}, P.,
  {Pool}, P., {Soufli}, R., {Windt}, D.~L., {Beardsley}, S., {Clapp}, M.,
  {Lang}, J., \& {Waltham}, N. 2012, \solphys, 275, 17

\bibitem[{{Mandrini} {et~al.}(1997){Mandrini}, {D\'emoulin}, {Bagala}, {van
  Driel-Gesztelyi}, {Henoux}, {Schmieder}, \& {Rovira}}]{Mandrini_al97}
{Mandrini}, C.~H., {D\'emoulin}, P., {Bagala}, L.~G., {van Driel-Gesztelyi},
  L., {Henoux}, J.~C., {Schmieder}, B., \& {Rovira}, M.~G. 1997, \solphys, 174,
  229

\bibitem[{{Masson} {et~al.}(2012){Masson}, {Aulanier}, {Pariat}, \&
  {Klein}}]{Masson_al12a}
{Masson}, S., {Aulanier}, G., {Pariat}, E., \& {Klein}, K.-L. 2012, \solphys,
  276, 199

\bibitem[{{Masson} {et~al.}(2009{\natexlab{a}}){Masson}, {Klein},
  {B{\"u}tikofer}, {Fl{\"u}ckiger}, {Kurt}, {Yushkov}, \&
  {Krucker}}]{Masson_al09a}
{Masson}, S., {Klein}, K., {B{\"u}tikofer}, R., {Fl{\"u}ckiger}, E., {Kurt},
  V., {Yushkov}, B., \& {Krucker}, S. 2009{\natexlab{a}}, \solphys, 257, 305

\bibitem[{{Masson} {et~al.}(2009{\natexlab{b}}){Masson}, {Pariat}, {Aulanier},
  \& {Schrijver}}]{Masson_al09b}
{Masson}, S., {Pariat}, E., {Aulanier}, G., \& {Schrijver}, C.~J.
  2009{\natexlab{b}}, \apj, 700, 559

\bibitem[{{Moreno-Insertis} {et~al.}(2008){Moreno-Insertis}, {Galsgaard}, \&
  {Ugarte-Urra}}]{Moreno_al08}
{Moreno-Insertis}, F., {Galsgaard}, K., \& {Ugarte-Urra}, I. 2008, \apjl, 673,
  L211

\bibitem[{{O'Dwyer} {et~al.}(2010){O'Dwyer}, {Del Zanna}, {Mason}, {Weber}, \&
  {Tripathi}}]{ODwyer_al10}
{O'Dwyer}, B., {Del Zanna}, G., {Mason}, H.~E., {Weber}, M.~A., \& {Tripathi},
  D. 2010, \aap, 521, A21

\bibitem[{{Pariat} {et~al.}(2009){Pariat}, {Antiochos}, \&
  {DeVore}}]{Pariat_al09}
{Pariat}, E., {Antiochos}, S.~K., \& {DeVore}, C.~R. 2009, \apj, 691, 61

\bibitem[{{Parnell} {et~al.}(2010){Parnell}, {Haynes}, \&
  {Galsgaard}}]{Parnell_al10}
{Parnell}, C.~E., {Haynes}, A.~L., \& {Galsgaard}, K. 2010, Journal of
  Geophysical Research (Space Physics), 115, 2102

\bibitem[{{Pesnell} {et~al.}(2012){Pesnell}, {Thompson}, \&
  {Chamberlin}}]{Pesnell_al12}
{Pesnell}, W.~D., {Thompson}, B.~J., \& {Chamberlin}, P.~C. 2012, \solphys,
  275, 3

\bibitem[{{Priest} \& {D{\'e}moulin}(1995)}]{PriestDemoulin95}
{Priest}, E.~R. \& {D{\'e}moulin}, P. 1995, \jgr, 100, 23443

\bibitem[{{Priest} {et~al.}(2003){Priest}, {Hornig}, \& {Pontin}}]{PriHorPon03}
{Priest}, E.~R., {Hornig}, G., \& {Pontin}, D.~I. 2003, \jgr, 108, 1285

\bibitem[{{Reeves} {et~al.}(2010){Reeves}, {Linker}, {Miki{\'c}}, \&
  {Forbes}}]{Reeves_al10}
{Reeves}, K.~K., {Linker}, J.~A., {Miki{\'c}}, Z., \& {Forbes}, T.~G. 2010,
  \apj, 721, 1547

\bibitem[{{Reeves} {et~al.}(2007){Reeves}, {Warren}, \& {Forbes}}]{Reeves_al07}
{Reeves}, K.~K., {Warren}, H.~P., \& {Forbes}, T.~G. 2007, \apj, 668, 1210

\bibitem[{{Reid} {et~al.}(2012){Reid}, {Vilmer}, {Aulanier}, \&
  {Pariat}}]{Reid_al12}
{Reid}, H.~A.~S., {Vilmer}, N., {Aulanier}, G., \& {Pariat}, E. 2012, \aap,
  547, A52

\bibitem[{{Rickard} \& {Titov}(1996)}]{RickardTitov96}
{Rickard}, G.~J. \& {Titov}, V.~S. 1996, \apj, 472, 840

\bibitem[{{Savcheva} {et~al.}(2012){Savcheva}, {Pariat}, {van Ballegooijen},
  {Aulanier}, \& {DeLuca}}]{Savcheva_al12}
{Savcheva}, A., {Pariat}, E., {van Ballegooijen}, A., {Aulanier}, G., \&
  {DeLuca}, E. 2012, \apj, 750, 15

\bibitem[{{Schou} {et~al.}(2012){Schou}, {Scherrer}, {Bush}, {Wachter},
  {Couvidat}, {Rabello-Soares}, {Bogart}, {Hoeksema}, {Liu}, {Duvall}, {Akin},
  {Allard}, {Miles}, {Rairden}, {Shine}, {Tarbell}, {Title}, {Wolfson},
  {Elmore}, {Norton}, \& {Tomczyk}}]{Schou_al12}
{Schou}, J., {Scherrer}, P.~H., {Bush}, R.~I., {Wachter}, R., {Couvidat}, S.,
  {Rabello-Soares}, M.~C., {Bogart}, R.~S., {Hoeksema}, J.~T., {Liu}, Y.,
  {Duvall}, T.~L., {Akin}, D.~J., {Allard}, B.~A., {Miles}, J.~W., {Rairden},
  R., {Shine}, R.~A., {Tarbell}, T.~D., {Title}, A.~M., {Wolfson}, C.~J.,
  {Elmore}, D.~F., {Norton}, A.~A., \& {Tomczyk}, S. 2012, \solphys, 275, 229

\bibitem[{{Schrijver} \& {De Rosa}(2003)}]{SchriDerosa03}
{Schrijver}, C.~J. \& {De Rosa}, M.~L. 2003, \solphys, 212, 165

\bibitem[{{Seaton} {et~al.}(2013){Seaton}, {De Groof}, {Shearer}, {Berghmans},
  \& {Nicula}}]{Seaton_al13}
{Seaton}, D.~B., {De Groof}, A., {Shearer}, P., {Berghmans}, D., \& {Nicula},
  B. 2013, ArXiv e-prints

\bibitem[{{Sheeley} {et~al.}(1997){Sheeley}, {Wang}, {Hawley}, {Brueckner},
  {Dere}, {Howard}, {Koomen}, {Korendyke}, {Michels}, {Paswaters}, {Socker},
  {St.~Cyr}, {Wang}, {Lamy}, {Llebaria}, {Schwenn}, {Simnett}, {Plunkett}, \&
  {Biesecker}}]{Sheeley_al97}
{Sheeley}, Jr., N.~R., {Wang}, Y.-M., {Hawley}, S.~H., {Brueckner}, G.~E.,
  {Dere}, K.~P., {Howard}, R.~A., {Koomen}, M.~J., {Korendyke}, C.~M.,
  {Michels}, D.~J., {Paswaters}, S.~E., {Socker}, D.~G., {St.~Cyr}, O.~C.,
  {Wang}, D., {Lamy}, P.~L., {Llebaria}, A., {Schwenn}, R., {Simnett}, G.~M.,
  {Plunkett}, S., \& {Biesecker}, D.~A. 1997, \apj, 484, 472

\bibitem[{{Sun} {et~al.}(2013){Sun}, {Hoeksema}, {Liu}, {Aulanier}, {Su},
  {Hannah}, \& {Hock}}]{Sun_al13}
{Sun}, X., {Hoeksema}, J.~T., {Liu}, Y., {Aulanier}, G., {Su}, Y., {Hannah},
  I.~G., \& {Hock}, R.~A. 2013, ArXiv e-prints

\bibitem[{{Titov} {et~al.}(2011){Titov}, {Miki{\'c}}, {Linker}, {Lionello}, \&
  {Antiochos}}]{Titov_al11}
{Titov}, V.~S., {Miki{\'c}}, Z., {Linker}, J.~A., {Lionello}, R., \&
  {Antiochos}, S.~K. 2011, \apj, 731, 111

\bibitem[{{Titov} {et~al.}(2012){Titov}, {Mikic}, {T{\"o}r{\"o}k}, {Linker}, \&
  {Panasenco}}]{Titov_al12}
{Titov}, V.~S., {Mikic}, Z., {T{\"o}r{\"o}k}, T., {Linker}, J.~A., \&
  {Panasenco}, O. 2012, \apj, 759, 70

\bibitem[{{van Driel-Gesztelyi} {et~al.}(2012){van Driel-Gesztelyi}, {Culhane},
  {Baker}, {D{\'e}moulin}, {Mandrini}, {DeRosa}, {Rouillard}, {Opitz},
  {Stenborg}, {Vourlidas}, \& {Brooks}}]{vanDriel_al12}
{van Driel-Gesztelyi}, L., {Culhane}, J.~L., {Baker}, D., {D{\'e}moulin}, P.,
  {Mandrini}, C.~H., {DeRosa}, M.~L., {Rouillard}, A.~P., {Opitz}, A.,
  {Stenborg}, G., {Vourlidas}, A., \& {Brooks}, D.~H. 2012, \solphys, 228

\bibitem[{{Wang} {et~al.}(2007){Wang}, {Biersteker}, {Sheeley}, {Koutchmy},
  {Mouette}, \& {Druckm{\"u}ller}}]{Wang_al07b}
{Wang}, Y., {Biersteker}, J.~B., {Sheeley}, Jr., N.~R., {Koutchmy}, S.,
  {Mouette}, J., \& {Druckm{\"u}ller}, M. 2007, \apj, 660, 882

\bibitem[{{Wang} {et~al.}(2012){Wang}, {Grappin}, {Robbrecht}, \&
  {Sheeley}}]{WangY_al12}
{Wang}, Y.-M., {Grappin}, R., {Robbrecht}, E., \& {Sheeley}, Jr., N.~R. 2012,
  \apj, 749, 182

\bibitem[{{Wang} {et~al.}(2000){Wang}, {Sheeley}, {Socker}, {Howard}, \&
  {Rich}}]{Wang_al00}
{Wang}, Y.-M., {Sheeley}, N.~R., {Socker}, D.~G., {Howard}, R.~A., \& {Rich},
  N.~B. 2000, \jgr, 105, 25133

\bibitem[{{Wang} \& {Sheeley}(1991)}]{WangSheeley91}
{Wang}, Y.-M. \& {Sheeley}, Jr., N.~R. 1991, \apjl, 372, L45

\bibitem[{{Wang} {et~al.}(1998){Wang}, {Sheeley}, {Walters}, {Brueckner},
  {Howard}, {Michels}, {Lamy}, {Schwenn}, \& {Simnett}}]{Wang_al98}
{Wang}, Y.-M., {Sheeley}, Jr., N.~R., {Walters}, J.~H., {Brueckner}, G.~E.,
  {Howard}, R.~A., {Michels}, D.~J., {Lamy}, P.~L., {Schwenn}, R., \&
  {Simnett}, G.~M. 1998, \apjl, 498, L165

\bibitem[{{Zhao} {et~al.}(2009){Zhao}, {Zurbuchen}, \& {Fisk}}]{Zhao_al09}
{Zhao}, L., {Zurbuchen}, T.~H., \& {Fisk}, L.~A. 2009, \grl, 36, 14104

\end{thebibliography}
\IfFileExists{\jobname.bbl}{} {\typeout{}
\typeout{***************************************************************}
\typeout{***************************************************************}
\typeout{** Please run "bibtex \jobname" to obtain the bibliography} 
\typeout{** and re-run "latex \jobname" twice to fix references} 
\typeout{***************************************************************}
\typeout{***************************************************************}
\typeout{}}

\end{document}